\RequirePackage[compress]{cite} 
\documentclass[final,1p,linenumbers,compress,hyperref]{elsarticle}

\makeatletter
\newcommand*{\rom}[1]{\expandafter\@slowromancap\romannumeral #1@}
\makeatother

\usepackage{lineno,hyperref}
\usepackage{multirow}
\usepackage[utf8]{inputenc}

\usepackage{amssymb}
\usepackage{amsmath}
\usepackage{graphicx,color}

\newcommand{\roots} {\mbox{$\sqrt{\textit{s}_{NN}}$}}
\newcommand{\GeVc} {\mbox{GeV/$\textit{c}$}}

\def  \vn         {\mbox{$\textit{v}_{n}$  }} 
\def  \first       {\mbox{$\textit{v}_{1}$  }}
\def  \second   {\mbox{$\textit{v}_{2}$  }}
\def  \third      {\mbox{$\textit{v}_{3}$  }}
\def  \fourth    {\mbox{$\textit{v}_{4}$  }}
\def  \fifth       {\mbox{$\textit{v}_{5}$  }}

\def  \higher    {\mbox{$\textit{v}_{n > 3}$  }}

\def  \etas      {\mbox{$\eta / \textit{s}$  }}

\journal{Phys. Lett. B}

\begin{document}
\begin{frontmatter}

\title{Investigation of the linear and mode-coupled flow harmonics in Au+Au collisions at \roots = 200~GeV}

\author{
J.~Adam$^{6}$,
L.~Adamczyk$^{2}$,
J.~R.~Adams$^{39}$,
J.~K.~Adkins$^{30}$,
G.~Agakishiev$^{28}$,
M.~M.~Aggarwal$^{41}$,
Z.~Ahammed$^{61}$,
I.~Alekseev$^{3,35}$,
D.~M.~Anderson$^{55}$,
A.~Aparin$^{28}$
% ...
\\
(STAR Collaboration)
}

% ...
\address{$^{2}$AGH University of Science and Technology, FPACS, Cracow 30-059, Poland}
\address{$^{3}$Alikhanov Institute for Theoretical and Experimental Physics NRC "Kurchatov Institute", Moscow 117218, Russia}
% ...
\address{$^{6}$Brookhaven National Laboratory, Upton, New York 11973}
% ...
\address{$^{28}$Joint Institute for Nuclear Research, Dubna 141 980, Russia}
% ...
\address{$^{30}$University of Kentucky, Lexington, Kentucky 40506-0055}
% ...
\address{$^{35}$National Research Nuclear University MEPhI, Moscow 115409, Russia}
% ...
\address{$^{39}$Ohio State University, Columbus, Ohio 43210}
% ...
\address{$^{41}$Panjab University, Chandigarh 160014, India}
% ...
\address{$^{55}$Texas A\&M University, College Station, Texas 77843}
% ...
\address{$^{61}$Variable Energy Cyclotron Centre, Kolkata 700064, India}
% ...


\begin{abstract}
Flow harmonics ($\textit{v}_{n}$) of the Fourier expansion for the azimuthal distributions of hadrons are commonly employed to quantify the azimuthal anisotropy of particle production relative to the collision symmetry planes. While lower order Fourier coefficients ($\textit{v}_{2}$ and $\textit{v}_{3}$) are more directly related to the corresponding eccentricities of the initial state, the higher-order flow harmonics ($\textit{v}_{n>3}$) can be induced by a mode-coupled response to the lower-order anisotropies, in addition to a linear response to the same-order anisotropies.
These higher-order flow harmonics and their linear and  mode-coupled contributions can be used to more precisely constrain the initial conditions and the transport properties of the medium in theoretical models.
The multiparticle azimuthal cumulant method is used to measure the linear and  mode-coupled contributions in the higher-order anisotropic flow, the  mode-coupled response coefficients, and the correlations of the event plane angles for charged particles as functions of  centrality and transverse momentum in Au+Au collisions at nucleon-nucleon center-of-mass energy \roots = 200~GeV. The results are compared to similar LHC measurements as well as to several viscous hydrodynamic calculations with varying initial conditions. 
\end{abstract}


\end{frontmatter}



\section{Introduction}

Experimental studies of heavy-ion collisions at the Relativistic Heavy Ion Collider (RHIC) indicate that a state of matter predicted by Quantum Chromodynamics (QCD), called Quark-Gluon Plasma (QGP), is formed in these collisions. Many of the ongoing studies are aimed at characterizing the transport properties (particularly,  the specific shear viscosity: the ratio of shear viscosity to entropy density $\eta / \textit{s}$) of the QGP.  The azimuthal anisotropy of particle production relative to the collision symmetry planes, known as anisotropic flow, is a key observable in many such studies because it displays the viscous hydrodynamic response to the initial spatial distribution created in the early stages of the collision ~\cite{Heinz:2001xi,Hirano:2005xf,Huovinen:2001cy,Hirano:2002ds,Romatschke:2007mq,Luzum:2011mm,Song:2010mg,Qian:2016fpi,Magdy:2017ohf,Magdy:2017kji,Schenke:2011tv,Teaney:2012ke,Gardim:2012yp,Lacey:2013eia}.

The  anisotropic flow can be characterized by the Fourier expansion~\cite{Poskanzer:1998yz} of the particle azimuthal angle ($\phi$) distributions,
\begin{eqnarray}
\label{eq:1-1}
\frac{dN}{d\phi} = \dfrac{\textit{N}}{2\pi}  \left(  1+2\sum_{n=1}\textit{V}_{n} e^{-i \textit{n} \phi} \right)  ,
\end{eqnarray}
where $\textit{V}_{n} = \vn e^{i \textit{n} \Psi_{n}}$ is the n-th complex anisotropic flow vector, \vn and $\Psi_{n}$ represent the vector magnitude and direction, respectively. The flow coefficient \first is commonly termed as directed flow, \second is the elliptic flow, and \third is the triangular flow.  Anisotropic flow studies of higher-order flow harmonics \higher~\cite{Magdy:2019ojv,Adam:2019woz,Magdy:2018itt,Adamczyk:2017ird,Magdy:2017kji,Adamczyk:2017hdl,Alver:2010gr, Chatrchyan:2013kba}, correlation between different flow harmonics~\cite{STAR:2018fpo,Adamczyk:2017hdl,Qiu:2011iv, Adare:2011tg, Aad:2014fla, Aad:2015lwa} and flow fluctuations~\cite{Magdy:2018itt,Alver:2008zza,Alver:2010rt, Ollitrault:2009ie} have led to a deeper understanding of the initial conditions ~\cite{Busza:2018rrf} and the properties of the matter created in heavy-ion collisions.

In the hydrodynamic models, anisotropic flow arises from the evolution of the medium in the presence of initial-state energy density anisotropies, characterized by the complex eccentricity vectors~\cite{Alver:2010dn,Petersen:2010cw,Lacey:2010hw,Teaney:2010vd,Qiu:2011iv}:
\begin{eqnarray}
\mathrm{\mathcal{E}_{n}  \equiv \varepsilon_{n} e^{i {\textit{n}} \Phi_{n} } \equiv 
  - \frac{\int d^2r_\perp\, \textit{r}^{n}\,e^{i {\textit{n}} \varphi}\, \rho_\textit{e}(r,\varphi)}
           {\int d^2r_\perp\, \textit{r}^{n}\,\rho_\textit{e}(r,\varphi)}}, ~(\textit{n} ~>~ 1),
\label{epsdef1}
\end{eqnarray}
where  $\rho_{\textit{e}}(r,\varphi)$ is the initial anisotropic density profile, $\mathrm{\varepsilon_{n} = {\left< \left| \mathcal{E}_{n} \right|^2 \right>}^{1/2}}$ represents the eccentricity vectors magnitude and $\mathrm{\Phi_{n}}$ denotes the azimuthal direction of the eccentricity vector~\cite{Teaney:2010vd,Bhalerao:2014xra,Yan:2015jma}. 

The elliptic and triangular flow harmonics are, to a reasonable approximation, linearly proportional to the initial-state anisotropies, $\varepsilon_{{{2}}}$ and $\varepsilon_{{{3}}}$, respectively~\cite{Song:2010mg, Niemi:2012aj,Gardim:2014tya, Fu:2015wba,Holopainen:2010gz,Qin:2010pf,Qiu:2011iv,Gale:2012rq,Liu:2018hjh}: 
\begin{eqnarray}
\label{eq:1-2}
v_{n} = k_{n} \varepsilon_{n},
\end{eqnarray}
where $\textit{k}_{n}$ is the proportionality factor that encodes the medium response, and is expected to be  sensitive to \etas and the system lifetime~\cite{Heinz:2013th}.  Therefore, the ratio $\vn / \varepsilon_{n}$ (for $\textit{n} = 2, 3$) could be used as a tool to probe \etas of the QGP~\cite{Adam:2019woz}. In contrast, the higher-order flow harmonics are expected to arise from a mode-coupled (nonlinear) response to the lower-order eccentricities, $\varepsilon_{2}$ and/or $\varepsilon_{3}$~\cite{Teaney:2012ke,Bhalerao:2014xra,Yan:2015jma} in addition to linear response to the same-order initial-state anisotropies~\cite{Gardim:2011xv}:

\begin{eqnarray}\label{eq:1-3}
V_{4}  &=&   V_{4}^{\rm L} +  V_{4}^{\rm mc}  ~=~  V_{4}^{\rm L} +  \chi_{4,22} V_{2} V_{2},\\
V_{5}  &=&   V_{5}^{\rm L} +  V_{5}^{\rm mc}  ~=~  V_{5}^{\rm L} +  \chi_{5,23} V_{2} V_{3},
\end{eqnarray}
where $\textit{V}^{L}_{n}$ and $\textit{V}^{\rm mc}_{n}$ represents the linear and the mode-coupled contributions to the flow vector $\textit{V}_{n}$ respectively. The $\chi_{4,22}$ and $\chi_{5,23}$ are the mode-coupled response coefficients which define the magnitude of the  $\textit{V}^{\rm mc}_{n > 3}$ measured with respect to the lower-order symmetry plane angle(s). 
Also, the mode-coupled contribution of $\textit{V}_{n}$ is expected to reflect the correlation between different order flow symmetry planes, $\Psi_{n}$, which could shed light on the initial stage dynamics~\cite{Bilandzic:2013kga, Bhalerao:2014xra, Aad:2015lwa, ALICE:2016kpq, STAR:2018fpo,Zhou:2016eiz, Qiu:2012uy,Teaney:2013dta, Niemi:2015qia, Zhou:2015eya}.

The \second and \third harmonics are sensitive to the respective influence of the initial-state eccentricity and the final-state viscous attenuation, which have proven difficult to disentangle. The mode-coupled coefficients show characteristically different dependencies on the viscous attenuation and the initial-state eccentricity~\cite{Liu:2018hjh}. Therefore,  they can be used in conjunction with measurements for the \second and \third harmonics to leverage additional unique constraints for initial-state models, as well as reliable extraction of  transport coefficients. 

In this paper we report new differential and integral measurements of \fourth and \fifth and their mode-coupled response coefficients, 
obtained with the two- and multiparticle cumulant methods described in Section~\ref{Sec:2}. Measurements of these quantities as functions of collision centrality and charged particle transverse momentum, $p_{T}$, in Au+Au collisions at \roots = 200~GeV, are reported in Section~\ref{Sec:3}. 
The presented results and conclusions are summarized in Section~\ref{Sec:4}.

\section{Experimental setup and analysis method}\label{Sec:2}
\subsection{Experimental setup}
The data reported in this analysis were collected with the STAR detector at RHIC using a minimum-bias trigger~\cite{Judd:2018zbg}  in 2011. Charged particle tracks, measured in pseudorapidity range $|\eta|<1.0$ and covering all azimuthal angles of the Time Projection Chamber (TPC)~\cite{Anderson:2003ur}, are used to reconstruct the collision vertices. Collision centrality is determined from the measured event-by-event multiplicity with the assistance of the Monte Carlo Glauber simulation~\cite{Alver:2008aq}. 
 Tracks included in the analysis are required to have a distance of closest approach to the primary vertex of less than 3~cm, and to have at least $15$ TPC space points used in their reconstruction. In order to remove track splitting, the ratio of the number of fit points to the maximum possible number of TPC fit points was required to be larger than~$0.52$.  Tracks used in this study are restricted to transverse momentum $0.2<p_{T}<4$~\GeVc. Events are chosen with vertex positions within $\pm 30$~cm from the TPC center (along the beam direction), and within $\pm 2$~cm in the radial direction relative to the center of the beam intersection.
Also, the absolute difference between the two $\textit{z}$-vertex positions defined by the TPC and Vertex Position Detector is required to be less than $3$ cm to decrease beam-induced background and pileup.

The systematic uncertainties associated with the measurements presented in this work are estimated by changing different parameters of the analysis and comparing the results with their baseline values.
The systematic uncertainty associated with the event selection is estimated by using more restrictive requirements for the vertex positions determined by the TPC along the beam direction ($-30$ to $0$~cm or $0$ to $30$~cm instead of the nominal value of $\pm 30$~cm). 
 The systematic uncertainty arising from track selection is evaluated by employing more strict requirements: (i)  Distance of Closest Approach (DCA) is changed to be less than 2~cm instead of the standard value of 3~cm, and (ii) number of TPC space points from more than $15$ points to more than $20$ points.
The systematic uncertainty associated with the nonflow effects,  due to Bose-Einstein correlations,  resonance decays  and the fragments of individual jets, is estimated by investigating the impact of a pseudorapidity gap, $\Delta\eta~=~\eta_{1}-\eta_{2}$, for the track pairs used in the measurements. Studies were performed for  $\Delta\eta$ values of 0.6, 0.7, and 1.0.

Table~\ref{tab:1} shows the systematic  uncertainties evaluated for this work. The overall systematic uncertainty was calculated by combining uncertainties from different sources in quadrature. In the ensuing figures, the overall systematic uncertainties (which do not include those from $\Delta\eta$ variation) are shown as  open boxes;  statistical uncertainties are shown as vertical lines. 

\begin{table}[h!]
\begin{center}
 \begin{tabular}{|c|c|c|}
 \hline
 Variations of Quantities        &          Minimum value        &                   Maximum value                      \\
 \hline
  Event                                 &              2\%                   &                      4\%                                   \\
 \hline 
 Track                                  &              3\%                   &                      6\%                                   \\
 \hline
 $\Delta\eta$                      &              3\%                   &                      8\%                                    \\
 \hline
\end{tabular} 
\caption{The contributions to the total systematic uncertainties from various  sources.}
\label{tab:1}
\end{center}
\end{table}

\subsection{Analysis method}

The two- and multiparticle cumulant techniques are used in this work. The framework for the cumulant method is described in Refs.~\cite{Bilandzic:2010jr,Bilandzic:2013kga}, which was extended to the case of subevents in Refs.~\cite{Gajdosova:2017fsc,Jia:2017hbm}. 
In this work, the two-  and multiparticle correlations were constructed using the two-subevents cumulant method~\cite{Jia:2017hbm}, with particle weights, e.g. weighted with the particles acceptance correction, and $\Delta\eta~ > 0.7$ separation between the subevents $\textit{A}$ and $\textit{B}$ (\textit{i.e.}, $\eta_{A}~ > 0.35$ and $\eta_{B}~ < -0.35$). The use of the two-subevents method helps to suppress the nonflow correlations. The two-  and multiparticle correlations are written as:
\begin{eqnarray}\label{eq:2-1}
v^{\rm Inclusive}_{k} &=&  \langle  \langle \cos (k (\varphi^{A}_{1} -  \varphi^{B}_{2} )) \rangle \rangle^{1/2},
\end{eqnarray}
\begin{eqnarray}\label{eq:2-2}
C_{k,nm}                  &=&   \langle \langle \cos ( k \varphi_{1}^{A} - n \varphi_{2}^{B} -  m \varphi_{3}^{B}) \rangle \rangle ,
\end{eqnarray}
\begin{eqnarray}\label{eq:2-3}
\langle v_{n}^{2} v_{m}^{2}  \rangle &=& \langle \langle \cos ( n \varphi^{A}_{1} + m \varphi^{A}_{2} -  n \varphi^{B}_{3} -  m \varphi^{B}_{4}) \rangle \rangle,
\end{eqnarray}
where $\langle \langle \, \rangle \rangle$ indicates the average over all particles in a single event and then the average over all events, $\textit{k} =\textit{n}+\textit{m}$, $\textit{n} = 2$, $\textit{m} = 2$ or $3$, and $\varphi_{i}$ is the azimuthal angle of the $\textit{i}$-th particle. 

Using  Eqs.~(\ref{eq:2-1})-(\ref{eq:2-3}), the mode-coupled contribution in higher-order anisotropic flow harmonics, \fourth and \fifth, can be expressed as~\cite{Yan:2015jma,Bhalerao:2013ina}:
\begin{eqnarray}\label{eq:2-4}
v_{4}^{\rm mc} &=&  \frac{C_{4,22}} {\sqrt{\langle \mathrm{v_2^2 v_2^2 }\rangle}}, \\ 
                                                 &\sim & \langle v_{4} \, \cos (4 \Psi_{4} - 2\Psi_{2} - 2\Psi_{2}) \rangle,  \nonumber \\
v_{5}^{\rm mc} &=& \frac{C_{5,23}} {\sqrt{\langle \mathrm{v_2^2 v_3^2 }\rangle}}, \\ 
                                                 &\sim & \langle v_{5} \, \cos (5 \Psi_{5} - 2\Psi_{2} - 3\Psi_{3}) \rangle, \nonumber
\end{eqnarray}
and the linear contribution to \fourth  and \fifth can be given as:
\begin{eqnarray}\label{eq:2-5}
v_{4}^{L}  &=& \sqrt{ (v^{\rm Inclusive}_{4})^{\,2} - (v^{\rm mc}_{4})^{\,2}  }, \\ \nonumber
v_{5}^{L}  &=& \sqrt{ (v^{\rm Inclusive}_{5})^{\,2} - (v^{\rm mc}_{5})^{\,2}  }.  
\end{eqnarray}

Equation (\ref{eq:2-5}) assumes that the linear and mode-coupled contributions in \fourth  and \fifth are independent~\cite{Yan:2015jma,Magdy:2020bhd}.
The ratios of the mode-coupled contribution to the inclusive \fourth and \fifth are expected to measure the correlations between different order flow symmetry planes~\cite{Acharya:2017zfg} and are expressed as $\rho_{4,22}$ and $\rho_{5,23}$, respectively. The $\rho_{4,22}$ and $\rho_{5,23}$ can be given as:
\begin{eqnarray}\label{eq:2-6}
\rho_{4,22} &=& \frac{v^{\rm mc}_{4}}{v^{\rm Inclusive}_{4}}  = \langle  \cos (4 \Psi_{4} - 2 \Psi_{2} - 2 \Psi_{2}) \rangle, \\
\rho_{5,23} &=& \frac{v^{\rm mc}_{5}}{v^{\rm Inclusive}_{5}}  = \langle  \cos (5 \Psi_{5} - 2 \Psi_{2} - 3 \Psi_{3}) \rangle.
\end{eqnarray}

The mode-coupled response coefficients, $\chi_{4,22}$ and $\chi_{5,23}$, which quantify the contributions of the mode-coupling to the the higher-order anisotropic flow harmonics, are defined as
\begin{eqnarray}\label{eq:2-7}
\chi_{4,22} &=& \frac{v^{\rm mc}_{4}} {\sqrt{\langle  v_{2}^{2} \, v_{2}^{2} \rangle}} \\
\label{eq:2-8}
\chi_{5,23} &=& \frac{v^{\rm mc}_{5}} {\sqrt{\langle  v_{2}^{2} \, v_{3}^{2} \rangle}}.
\end{eqnarray}
In Eq.(\ref{eq:2-8}) for the differential $\chi_{5,23}$, this work further makes the approximation $\langle \textit{v}_{2}^{2} \textit{v}_{3}^{2}\rangle$ $\sim$  $\langle \textit{v}_{2}^{2} \rangle$ $\langle \textit{v}_{3}^{2} \rangle$~\cite{Bhalerao:2014xra}. These dimensionless ratios that represent the mode-coupled coefficients in Eq.(\ref{eq:1-3}) are expected to be weakly sensitive to viscous effects~\cite{Liu:2018hjh}.

\section{Results and discussion}\label{Sec:3}
\begin{figure*}[h!] 
  \vskip -0.2cm
 \centering{
 \includegraphics[width=0.9\linewidth, angle=0]{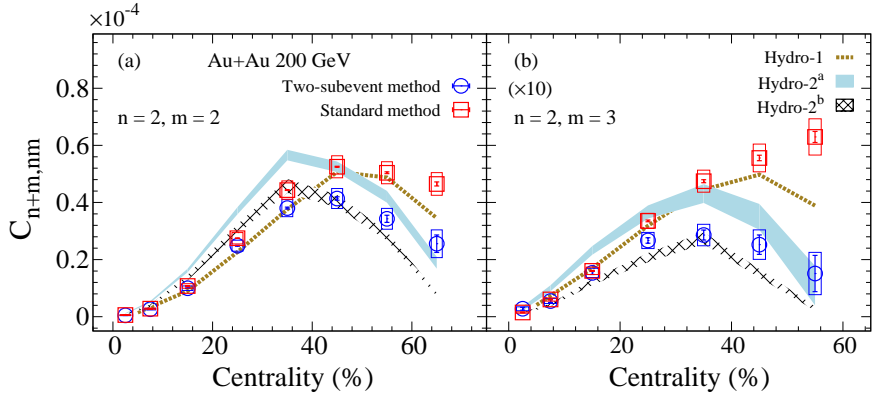}
  \vskip -0.3cm
\caption{Comparison of the $p_T$-integrated three-particle correlators, $C_{4,22}$ and $C_{5,23}$, for Au+Au collisions at \roots = 200~GeV, obtained with the standard (red squares) and the two-subevents cumulant (blue circles) methods. The respective systematic uncertainties, that do not include the nonflow contributions, are shown as open boxes. The vertical lines represent the statistical errors. The shaded bands indicate hydrodynamic model predictions Hydro$-1$~\cite{Alba:2017hhe}, Hydro$-2^{a}$ and Hydro$-2^{b}$~\cite{Schenke:2019ruo}.
\label{Fig:1} }}
 \vskip -0.1cm
\end{figure*}

In A+A collisions, short-range nonflow correlations contribute to the measured three-particle 
correlators $C_{4,22}$ and $C_{5,23}$~\cite{Magdy:2020bhd}.  However, such correlations can be 
reduced by using subevents cumulant methods~\cite{Jia:2017hbm}. Figure~\ref{Fig:1} compares 
the $C_{4,22}$ and $C_{5,23}$ values obtained from the standard (\textit{i.e.}, the three particles are selected using the entire detector acceptance) and the two-subevents cumulant methods as a function of centrality in the range  $0.2 < p_{T} <4.0$~\GeVc~ for Au+Au collisions at \roots = 200~GeV. The magnitudes of the measured $C_{4,22}$ and $C_{5,23}$ from the standard cumulant method are larger than those from the subevents cumulant method, compatible with the expectation that the subevents cumulant method can further reduce the nonflow correlations.
The shaded bands in Fig.~\ref{Fig:1}  indicate viscous hydrodynamic model 
predictions~\cite{Alba:2017hhe,Schenke:2019ruo}, as summarized in Table~\ref{tab:2}. 
Note that these model predictions include an  influence from changes in the initial- and final-state 
assumptions incorporated in model calculations. The model predictions, which were 
generated with the standard cumulant method, show good qualitative agreement with
both $C_{4,22}$ and $C_{5,23}$. However,  Hydro$-2^{b}$ with no hadronic cascade 
gives a better description of the data for $C_{4,22}$ and $C_{5,23}$ obtained with the two-subevents cumulant method.

\begin{table}[h!]
\begin{center}
 \begin{tabular}{|c|c|c|}
 \hline
 $   $                         & Hydro$-1$~\cite{Alba:2017hhe} &  Hydro$-2^{a/b}$~\cite{Schenke:2019ruo} \\
 \hline
 $\etas$                    &              0.05                             &                      0.12                                      \\
 \hline
 Initial conditions         &         TRENTO  Initial conditions    &          IP-Glasma  Initial conditions                 \\
 \hline
  Contributions            &         Hydro + Direct decays         &          (a)  Hydro + Hadronic cascade            \\
                                  &                                                  &          (b)  Hydro  only                                  \\
 \hline
\end{tabular} 
\caption{Summary description of the hydrodynamic simulations, Hydro$-1$~\cite{Alba:2017hhe}, and Hydro$-2^{a/b}$~\cite{Schenke:2019ruo}.
\label{tab:2}}
\end{center}
\end{table}
 \begin{figure*}[h!]
  \vskip -0.4cm
 \centering{
\includegraphics[width=0.9\linewidth,angle=0]{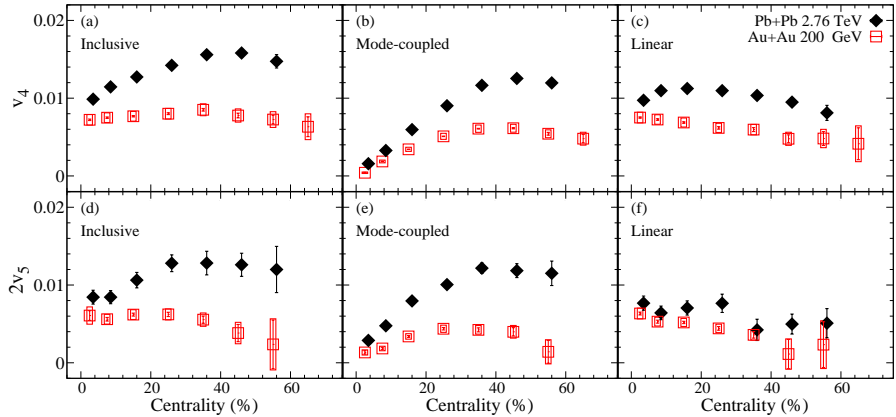}
\vskip -0.5cm
\caption{Comparison of the inclusive mode-coupled and linear higher-order flow harmonics \fourth and \fifth  obtained with the two-subevents cumulant method, as a function of centrality in the $p_{T}$ range $0.2 - 4.0$~\GeVc~ for Au+Au collisions at \roots = 200~GeV. The systematic uncertainties, that do not include the nonflow contributions, are shown as open boxes. 
The solid diamonds indicate LHC  measurements for the $p_{T}$ range from $0.2 -  5.0$~\GeVc~ for Pb+Pb collisions at \roots = 2.76~TeV~\cite{Acharya:2017zfg}.
 \label{Fig:2} }}
 \vskip -0.3cm
\end{figure*}
 \begin{figure*}[h!]
 \centering{
  \vskip -0.2cm
 \includegraphics[width=0.7\linewidth, angle=0]{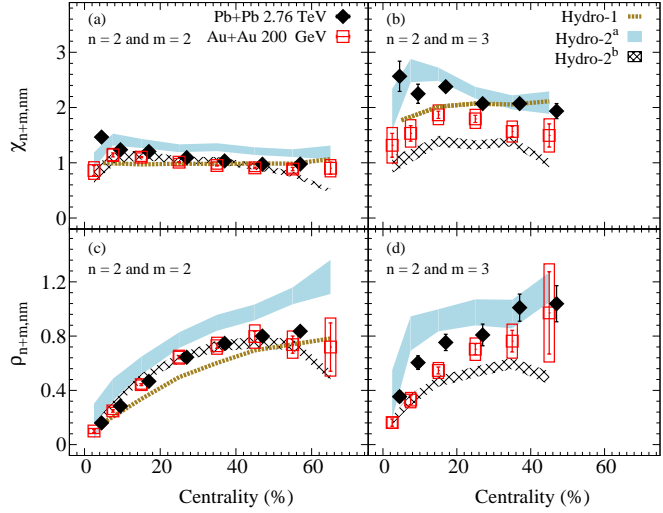}
 \vskip -0.5cm
 \caption{
 Results as a function of centrality in the $p_{T}$ range from $0.2$ to $4.0$~\GeVc~ for Au+Au collisions at \roots = 200~GeV. 
Panels (a) and (b) shows the mode-coupled response coefficients, $\chi_{4,22}$ and $\chi_{5,23}$, and panels (c) and (d) 
show the correlations of event plane angles, $\rho_{4,22}$ and $\rho_{5,23}$.  The results were obtained with the 
two-subevents cumulant method; the open boxes indicate the systematic uncertainties. 
The closed-symbols represents similar LHC  measurements in the $p_{T}$ range from $0.2$ to $5.0$~\GeVc~ for Pb+Pb collisions at \roots = 2.76~TeV~\cite{Acharya:2017zfg}. The shaded bands indicate hydrodynamic model predictions Hydro$-1$~\cite{Alba:2017hhe}, Hydro$-2^{a}$ and Hydro$-2^{b}$~\cite{Schenke:2019ruo}.
 \label{Fig:3} }}
 \vskip -0.4cm
 \end{figure*}
 \begin{figure*}[h!]
  \vskip -0.2cm
 \centering{
 \includegraphics[width=0.7\linewidth, angle=0]{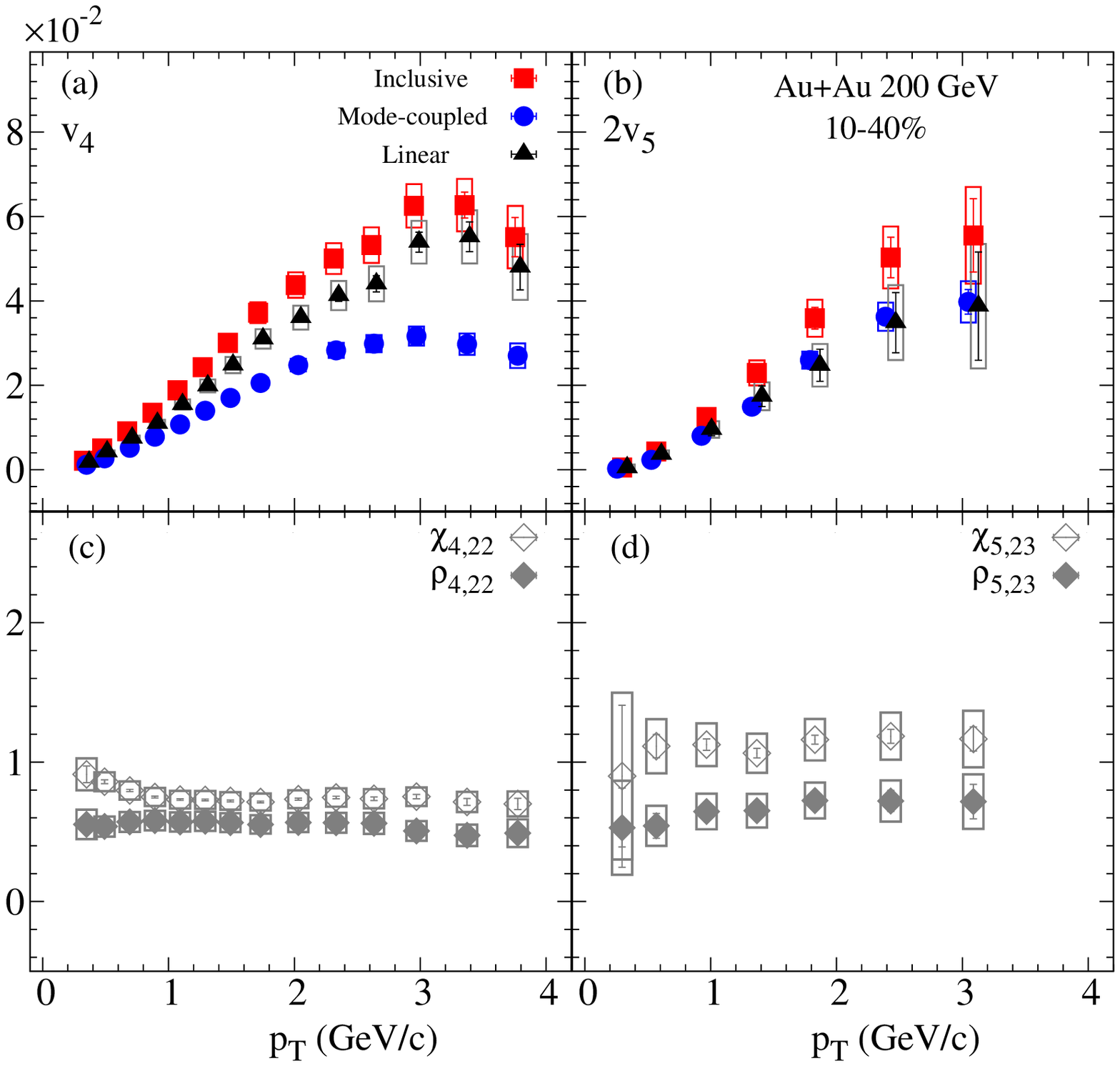}
 \vskip -0.5cm
 \caption{
 Results as a function of $p_{T}$ for 10-40\% central Au+Au collisions at \roots = 200~GeV. Panels (a) and (b) present the inclusive, linear and mode-coupled higher-order flow harmonics \fourth and \fifth  obtained with the two-subevents cumulant method. 
Panel (c)  presents  the $\chi_{4,22}$ and $\rho_{4,22}$,  while panel (d) presents the  $\chi_{5,23}$ and $\rho_{5,23}$. The open boxes indicate the systematic uncertainties.
 \label{Fig:4} }}
 \vskip -0.4cm
 \end{figure*}

The centrality dependence of the inclusive, linear and mode-coupled \fourth and \fifth in the $p_{T}$ range from $0.2$ to $4.0$~\GeVc~ for Au+Au collisions at \roots = 200~GeV are shown in Fig.~\ref{Fig:2}. They indicate that the linear mode of \fourth and \fifth  depends weakly on the collision centrality and constitutes the dominant contribution to the inclusive  \fourth and \fifth  in central collisions. 
These results are compared to similar LHC measurements in the $p_{T}$ range from $0.2$ to $5.0$~\GeVc~ and pseudorapidity range $|\eta|<0.8$ for Pb+Pb collisions at \roots = 2.76~TeV~\cite{Acharya:2017zfg}. The comparison indicates strikingly similar 
patterns for the RHIC and LHC measurements, albeit with a difference in the magnitude of the measurements.
This observed difference could result from a sizable difference in 
the $\left<p_T\right>$ for the $p_T$-integrated \fourth and \fifth measurements at RHIC and the LHC, respectively. 
Here, it is noteworthy that even though the $p_T$ range for both measurements is similar, the inverse slopes of 
the hadron $p_T$ spectra are larger at the LHC than at RHIC. Subtleties related to a difference in the  
viscous properties of the medium created at RHIC and LHC energies could also contribute to the observed difference in 
the magnitude of the measurements~\cite{Alba:2017hhe}.

The centrality dependence of the mode-coupled response coefficients, $\chi_{4,22}$ and $\chi_{5,23}$, for Au+Au collisions, is 
presented in Fig.~\ref{Fig:3}(a) and (b) for the range $0.2 < p_T < 4.0$~\GeVc. 
They show a weak centrality dependence, akin to the patterns  observed for similar measurements 
at the LHC for Pb+Pb collisions at \roots =~2.76 TeV~\cite{Acharya:2017zfg} (closed symbols).
These patterns suggests  that (i) the centrality dependence observed for  
the mode-coupled \fourth and \fifth (\textit{cf.}, Figs.~\ref{Fig:2}(b) and (e)) stems from the lower-order flow harmonics
and (ii)  the mode-coupled response coefficients are dominated by initial-state eccentricity couplings which have  
a weak dependence on beam energy.
The shaded bands in Figs.~\ref{Fig:3}(a) and (b) show that the 
predictions from the viscous hydrodynamic models~\cite{Alba:2017hhe,Schenke:2019ruo} summarized 
in Table~\ref{tab:2},  give a good qualitatively description of the $\chi_{4,22}$ and $\chi_{5,23}$ data. 
However, the predictions from Hydro$-1$ and Hydro$-2^{b}$ (cf. Table~\ref{tab:2}),  
give the overall closest description to $\chi_{4,22}$ and $\chi_{5,23}$.

Figures~\ref{Fig:3}(c) and (d) show the centrality dependence of the correlations of the event plane angles, 
$\rho_{4,22}$ and $\rho_{5,23}$, for $0.2 < p_{T} < 4.0$~\GeVc~ in Au+Au collisions at \roots = 200~GeV. 
The data suggest stronger event plane correlations in peripheral than in central collisions.
This centrality dependent pattern is also captured by the viscous hydrodynamic model 
predictions~\cite{Alba:2017hhe,Schenke:2019ruo} indicated by the shaded bands in the figure. 
The LHC $\rho_{4,22}$ and $\rho_{5,23}$ measurements for Pb+Pb collisions at \roots = 2.76~TeV~\cite{Acharya:2017zfg} (closed symbols), also indicate magnitudes and trends similar to those for the Au+Au collisions. This observation could be 
an indication that the correlation of event plane angles are dominated by initial-state effects.
 
The $p_T$ dependence of the inclusive, linear and mode-coupled higher-order flow harmonics, \fourth and \fifth, 
for 10-40\% central Au+Au collisions, are compared in Figs.~\ref{Fig:4}(a) and (b). They show that the $p_T$-dependent 
trends of the linear and mode-coupled contributions are similar to the inclusive \fourth and \fifth, as previously 
measured by the STAR collaboration~\cite{Magdy:2017kji,Adamczyk:2017ird}. This observation suggests that 
the linear and  mode-coupled contributions are driven by the same $p_T$-dependent physics processes. 
 The corresponding mode-coupled response coefficients $\chi_{4,22}$ and $\chi_{5,23}$ and the correlations 
of event plane angles $\rho_{4,22}$ and $\rho_{5,23}$ are shown in Figs.~\ref{Fig:4} (c) and (d). 
They indicate little,  if any,  $p_{T}$ dependence for the centrality selection presented. 
 These trends suggest that both dimensionless coefficients  are dominated by initial-state effects.

\section{Summary}\label{Sec:4}

In summary, we have presented new differential measurements of the charge-inclusive, linear and mode-coupled 
contributions to the higher-order anisotropic flow coefficients \fourth and \fifth, mode-coupled response 
coefficients $\chi_{4,22}$ and $\chi_{5,23}$ and the correlations of the event plane angles $\rho_{4,22}$ 
and $\rho_{5,23}$, for  Au+Au collisions at \roots = 200~GeV. 
The $p_T$-integrated measurements indicate a sizable centrality dependence for the mode-coupled contributions of \fourth and \fifth, whereas the linear contributions, that dominate the central collisions, show a weak centrality dependence. The \fourth and \fifth results are compared with similar LHC measurements which show larger  magnitude that could be driven by the difference in the viscous effects and the mean $p_{T}$ between RHIC and LHC energies.
The $\chi_{4,22}$ and $\chi_{5,23}$  show a weak centrality dependence, however the $\rho_{4,22}$ and $\rho_{5,23}$ increase from central to peripheral collisions.  These dimensionless  coefficients show magnitudes and trends which are similar to 
those observed for  LHC measurements, suggesting that the correlations of event plane angles as well as the mode-coupled response coefficients are dominated by initial-state effects. This is further supported by the observed $p_{T}$ independence 
of the $\chi_{4,22}$, $\chi_{5,23}$, $\rho_{4,22}$ and $\rho_{5,23}$.
Viscous hydrodynamic model comparisons to the data indicate good qualitatively agreement. 
However, none of the models provide a simultaneous description of the three-particle correlations, 
the mode-coupled response coefficients, and the correlations of event plane angles. 
These higher-order flow measurements could provide additional stringent constraints 
to discern between initial state models and aid precision extraction of the  transport properties of the
medium produced in the collisions.

\section*{Acknowledgments}
%
We thank the RHIC Operations Group and RCF at BNL, the NERSC Center at LBNL, and the Open Science Grid consortium for providing resources and support.  This work was supported in part by the Office of Nuclear Physics within the U.S. DOE Office of Science, the U.S. National Science Foundation, the Ministry of Education and Science of the Russian Federation, National Natural Science Foundation of China, Chinese Academy of Science, the Ministry of Science and Technology of China and the Chinese Ministry of Education, the Higher Education Sprout Project by Ministry of Education at NCKU, the National Research Foundation of Korea, Czech Science Foundation and Ministry of Education, Youth and Sports of the Czech Republic, Hungarian National Research, Development and Innovation Office, New National Excellency Programme of the Hungarian Ministry of Human Capacities, Department of Atomic Energy and Department of Science and Technology of the Government of India, the National Science Centre of Poland, the Ministry  of Science, Education and Sports of the Republic of Croatia, RosAtom of Russia and German Bundesministerium fur Bildung, Wissenschaft, Forschung and Technologie (BMBF), Helmholtz Association, Ministry of Education, Culture, Sports, Science, and Technology (MEXT) and Japan Society for the Promotion of Science (JSPS).

\section*{References}

\bibliographystyle{elsarticle-num}
\bibliography{ref}

%


\end{document}